\documentclass[floatfix,aps,prl,notitlepage,amsmath,amssymb,superscriptaddress,10pt,nobibnotes]{revtex4-1}
\pdfoutput=1

\usepackage[english]{babel}
\usepackage{amsfonts, amssymb, amsmath, dsfont,mathrsfs}
\usepackage{graphicx}
\usepackage{color}
\usepackage[utf8]{inputenc}
\usepackage[T1]{fontenc}
\usepackage[svgnames]{xcolor}
\usepackage{textcomp}
\usepackage{siunitx}
\usepackage{multibib}
\newcites{supp}{Supplementary References}

\usepackage[left=3cm,right=3cm,top=3.5cm,bottom=3.5cm]{geometry} 

\usepackage{hyperref}
\hypersetup{colorlinks,urlcolor=DarkBlue,linkcolor=DarkBlue,citecolor=DarkBlue,pdfdisplaydoctitle=true,pdfpagemode=UseOutlines,bookmarksnumbered=true,bookmarksopen=true} 

\makeatletter
\renewcommand{\eqref}[1]{\textup{\tagform@{\ref*{#1}}}}
\makeatother

\makeatletter
\renewcommand{\figurename}{Fig.}
\renewcommand*{\fnum@figure}{{\normalfont\bfseries \figurename~\thefigure}}
\makeatother

\renewcommand{\bm}[1]{\boldsymbol{\mathbf{#1}}} 

\newcommand{\beginsupplement}{%
        \setcounter{table}{0}
        \renewcommand{\thetable}{S\arabic{table}}%
        \setcounter{figure}{0}
        \renewcommand{\thefigure}{S\arabic{figure}}%
        \setcounter{section}{0}
        \renewcommand{\thesection}{\Roman{section}}
        \setcounter{equation}{0}
        \renewcommand{\theequation}{S\arabic{equation}}
     }

\usepackage{titlesec}
\titleformat{\section}
  {\normalfont\large\bfseries}{\thesection}{1em}{}
\titleformat{\subsection}
  {\normalfont\normalsize\bfseries}{\thesubsection}{1em}{}

\begin{document}

\title{Noise-robust latent vector reconstruction in ptychography using deep generative models}

\author{Jacob Seifert}
\altaffiliation{\href{mailto:j.seifert@uu.nl}{j.seifert@uu.nl}}
\affiliation{Nanophotonics, Debye Institute for Nanomaterials Science and Centre for Extreme Matter and Emergent Phenomena, Utrecht University, P.O. Box 80000, 3508 TA Utrecht, The Netherlands}
\author{Yifeng Shao}
\affiliation{Imaging Physics Department, Applied Science Faculty, Delft University of Technology, The Netherlands}
\author{Allard P.\ Mosk}
\affiliation{Nanophotonics, Debye Institute for Nanomaterials Science and Centre for Extreme Matter and Emergent Phenomena, Utrecht University, P.O. Box 80000, 3508 TA Utrecht, The Netherlands}

\begin{abstract}
\noindent Computational imaging is increasingly vital for a broad spectrum of applications, ranging from biological to material sciences. 
This includes applications where the object is known and sufficiently sparse, allowing it to be described with a reduced number of parameters. When no explicit parameterization is available, a deep generative model can be trained to represent an object in a low-dimensional latent space. 
In this paper, we harness this dimensionality reduction capability of autoencoders to search for the object solution within the latent space rather than the object space. 
We demonstrate a novel approach to ptychographic image reconstruction by integrating a deep generative model obtained from a pre-trained autoencoder within an Automatic Differentiation Ptychography (ADP) framework. 
This approach enables the retrieval of objects from highly ill-posed diffraction patterns, offering an effective method for noise-robust latent vector reconstruction in ptychography. 
Moreover, the mapping into a low-dimensional latent space allows us to visualize the optimization landscape, which provides insight into the convexity and convergence behavior of the inverse problem.
With this work, we aim to facilitate new applications for sparse computational imaging such as when low radiation doses or rapid reconstructions are essential.
\end{abstract}

\maketitle


\section{Introduction}
Obtaining clear and accurate images under noisy conditions is paramount in many scientific and industrial applications. 
Whether in medical diagnostics, materials analysis, or semiconductor inspection, noise-robust imaging techniques can mean the difference between precise understanding and potential misinterpretation. 
Typical noise removal techniques, like median filtering~\cite{Chang2009-tw}, anisotropic diffusion~\cite{Perona1990, WANG20131729} and BM3D~\cite{Dabov}, have been shown to be effective and applicable across various domains, but are usually limited to image postprocessing and restoration.
In the field of computational imaging, where an image is algorithmically recovered from noisy intensity measurements, an interesting option becomes available: noise-robustness can be intrinsically included in the process of solving the inverse problem. 
For example, techniques such as accurate modeling of the underlying noise statistics~\cite{Thibault2012-mn, PhysRevA.102.043516}, sparse modeling~\cite{Katkovnik2013-ms, KATKOVNIK201772, Schloz:20}, deep denoiser priors~\cite{Goy2018-ie, Aslan2021-kw, Chen2022-qf}, and regularization by denoising~\cite{venkatakrishnan2013plug, Romano2017-rn, Metzler2018-ei, reehorst2018regularization, Wu2019-by} have been explored in this context.

Ptychography, a computational imaging method, has seen exponential growth in the number of related publications in recent years~\cite{Wang2023-ca}. 
The origin of ptychography dates back to Hoppe's 1969 work, where they introduced a method for phase retrieval from electron diffraction interference~\cite{Hoppe1969-wz}.
This foundational concept was refined and first named "ptychography" in a subsequent paper the following year~\cite{Hegerl1970-tz}. 
The process involves illuminating a thin object with a localized and coherent beam.
The illumination field is diffracted by the object and propagates in free space to form a diffraction pattern on a camera sensor~\cite{Rodenburg2019-kw}. 
By laterally translating the object or illumination field to overlapping regions, the object's phase and amplitude can be retrieved using iterative algorithms~\cite{Rodenburg2004-oi}. 
Variations of this method include Fourier ptychography (FP), which uses a microscope objective to collect diffracted light and computationally synthesizes an image in the spatial domain~\cite{Zheng2013-cg, Konda2020-jk}.
Interestingly, reciprocity relations allow for the conversion of acquired data between both modalities, opening doors for the integration and mutual enhancement of various reconstruction algorithms~\cite{Loetgering2023-oh}. 
In the past decade, numerous algorithmic extensions have been developed to leverage redundancy in ptychographic measurements, enhancing image quality by recovering parameters such as the illumination field~\cite{Maiden2009-pn, Du2020-pd, Du2023-ib}, scanning position errors~\cite{Maiden2012-xo, Dwivedi2018-yy}, object-camera distance~\cite{Loetgering2020-ug}, and multiple incoherent modes~\cite{Thibault2013-kg, Li2016-fe}.

The inherent data redundancy in ptychography also provides a unique opportunity for data-driven reconstruction techniques such as Automatic Differentiation Ptychography (ADP), which aims to optimize a loss function using gradient-descent and differentiable modeling~\cite{Nashed2017-rf, Ghosh2018-jf, Kandel2019-da, Du2020-sh}. 
This loss function is derived from the intensity prediction of a physics-based forward model, actual data, and regularization terms. 
Utilizing automatic differentiation (AD), this approach enables the simultaneous and joint reconstruction of multiple relevant parameters~\cite{Seifert2021-fr}. 
ADP enhances the reconstruction process by offering portability, flexibility, and adaptability to changes in the forward model. This includes fusing multiple camera sensors~\cite{Maathuis2022-jl}, adjusting the loss function for mixed noise statistics~\cite{seifert2023maximumlikelihood}, or tailoring a maximum-information illumination scheme~\cite{Bouchet2021-mb}.
An intriguing extension of data-driven image retrieval involves integrating deep neural networks (DNNs) with the reconstruction algorithm. 
One approach is combining physics knowledge and machine learning to achieve optimal experimental designs~\cite{chakrabarti2016learning, kellman2019physics, metzler2020deep}. 
Another approach is end-to-end mapping, where DNNs learn a direct mapping between the object and diffraction image domains~\cite{Sinha2017-op, Kappeler2017-lo, Jin2017, Sun2018-bg, boominathan2018phase, cherukara2020real, Li:22, Ye:22}. 
This method reduces the computationally slow and costly reconstruction process, even enabling real-time inference in some cases~\cite{Cherukara2018-au, Metzler:20, babu2023}. However, it requires a large amount of training data.
While the training data can be generated numerically in simulation, Sinha \textit{et al.}~\cite{Sinha2017-op} acquire them \textit{in situ} by projecting thousands of phase objects onto an SLM, whereas in Cherukara \textit{et al.}~\cite{cherukara2020real}, the training data is obtained from iterative phase retrieval of experimental data. 
Although these methods allow the model to learn physical system inaccuracies, they are time-consuming and not portable to different optical systems.
Given that wave propagation is well-described by the Helmholtz equation, requiring a DNN to learn this may be unnecessary. This insight leads to physics-informed and deep learning (DL) assisted computational imaging. For instance, Goy \textit{et al.}\cite{Goy2018-ie} found that physics-informed phase retrieval outperformed end-to-end methods for low photon counts. 
Similarly, Metzler \textit{et al.}~\cite{Metzler2018-ei} utilize a convolutional neural network (CNN) known as DnCNN~\cite{zhang2017beyond} as a denoising regularizer, improving the reconstruction quality by exploiting the natural absence of additive Gaussian noise in images, while Chang \textit{et al.}~\cite{Chang2023-yk} advance this approach by incorporating a complex-domain neural network that leverages the latent correlations between amplitude and phase for improved coherent imaging reconstructions.
A recent development in this field is the realization that the structure of deep generative networks can capture image statistics even before learning~\cite{dmitry2020deep}, improving ptychography reconstruction quality under the concept of deep image priors without requiring a preceding training procedure~\cite{du2021using, Chen2022-qf, barutcu2022compressive}.

In this paper, we introduce a method that combines a fully physics-based ptychography reconstruction framework with a pre-trained deep generative model. 
When prior knowledge indicates that a sample is sparse in an unknown basis, we show that the learned low-dimensional representation in latent space enables accurate image reconstruction even under extremely challenging noise conditions.
The integration of the deep generative model serves two distinct functions.
First, in a pre-training step, an under-complete autoencoder~\cite{Goodfellow2016-oh} learns an implicit parameterization of images belonging to a specific class (e.g., MNIST~\cite{lecun1998mnist}).
Then, this learned model is utilized to successfully reconstruct images from ill-posed diffraction data.
We empirically demonstrate noise-robust latent vector reconstruction using experimental data from a photolithographically manufactured sample. 
The compact space spanned by the latent vectors allows us to visualize and study the optimization landscape in ptychography through principal component analysis, a novel approach to the best of our knowledge.
Lastly, we quantify the reconstruction quality as a function of the total number of photons in the illumination field through numerical simulations. We find that latent vector reconstruction begins to produce faithful reconstructions with an average of 0.001 photons per camera pixel, even in the presence of readout noise. However, as the total number of photons approaches the degrees of freedom in conventional reconstruction, the performance of our approach diminishes in comparison due to its inability to produce comparably high-definition images.
We provide the raw ptychography data and an open-source implementation of ptychographic latent vector reconstruction underlying this paper in~\cite{Seifert2023-bk}.

\section{Methods}
\subsection{Optical Setup}

\begin{figure*}[htb]
\centering
\includegraphics[width=0.84\textwidth]{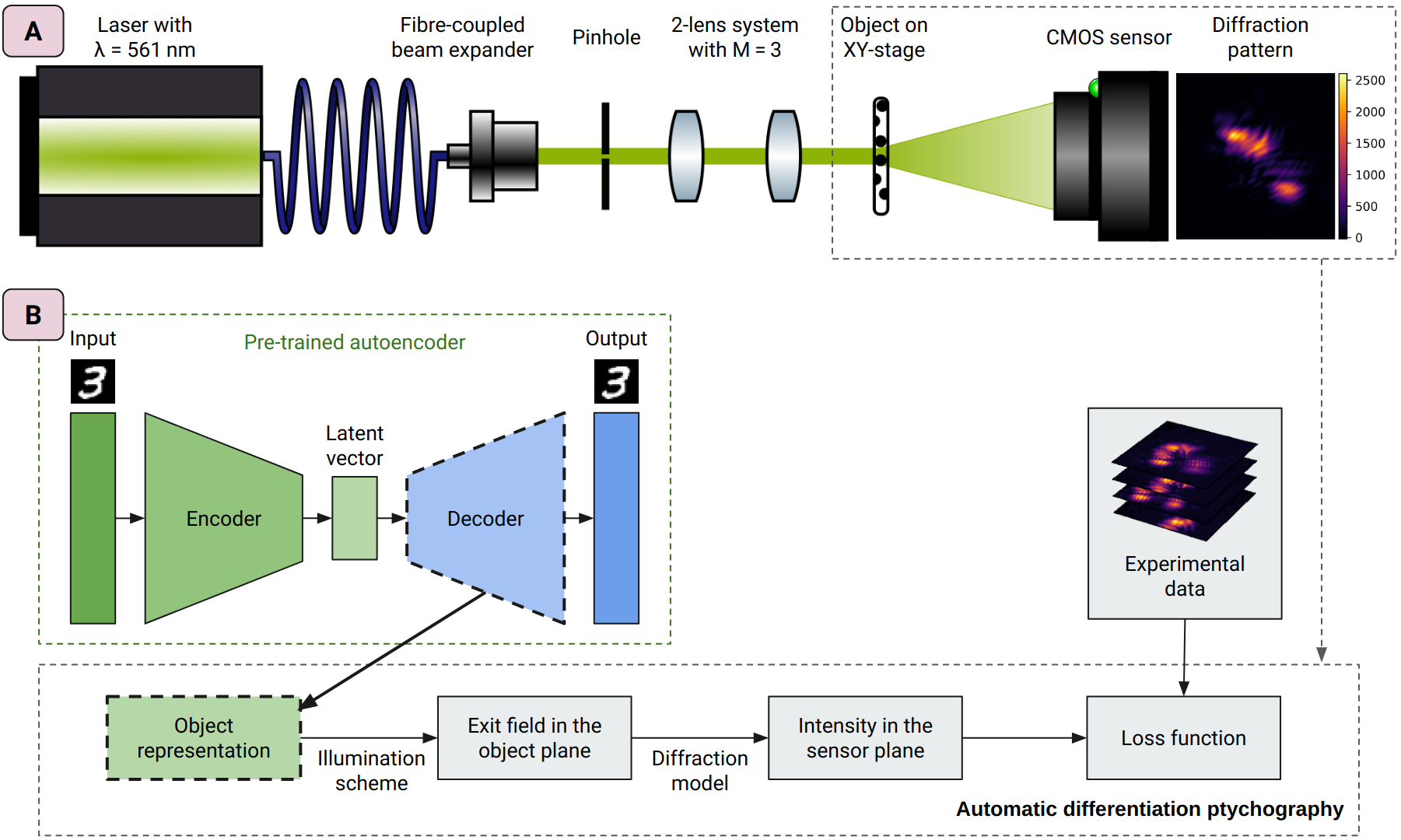}
\caption{(A) Schematic of the optical setup used for ptychography. A 500-µm pinhole is illuminated with coherent light at $\lambda = \SI{561}{nm}$ and relayed onto the object using a 2-lens system. The object is moved laterally through the beam using a computer-controlled XY stage, and a CMOS camera sensor records the diffraction intensities \SI{6.5}{cm} downstream from the object. (B) Diagram of the Automatic Differentiation Ptychography (ADP) framework, which models the physical system beginning from the object illumination. In the conventional mode, the object is represented by complex-valued pixels. With a pre-trained autoencoder for a specific class of objects, the decoder can be integrated into the ADP framework as a deep generative model, allowing the object to be represented as a latent vector and significantly reducing the number of free parameters.}
\label{setup}
\end{figure*}

We employ a ptychography setup in transmission geometry as illustrated in Fig. \ref{setup}A.
A continuous-wave laser (Cobolt Jive 100\texttrademark) with a wavelength of $\lambda = \SI{561}{nm}$ is coupled into a polarization-maintaining single-mode fiber. 
Then, a fiber-coupled collimator (60FC-L-0-M75-26, Schäfter+Kirchhoff) expands the beam to \SI{25}{mm} in diameter, illuminating a 500-µm pinhole. 
This pinhole is imaged onto the object using a 2-lens system with a magnification of $M = 3$, resulting in a circular illumination field with a uniform phase at the object plane.
The object, a binary hand-drawn digit, is laterally scanned through the beam using stepper motor actuators (ZFS25B, Thorlabs).
Diffraction patterns are recorded \SI{6.5}{cm} downstream of the object using a CMOS camera sensor (acA2440-35um, Basler) with a pixel size of \SI{3.45}{\micro\meter} and $1024 \times 1024$ total pixels.
For calibration of the illumination field, object-camera distance, and actuator position inaccuracies, we employ a Fermat spiral scanning pattern~\cite{Huang2014-yo} with 96 positions and an overlap of \SI{80}{\percent}. 
At 16 uniformly distributed positions during this calibration scan, we capture an additional evaluation series of diffraction patterns at a lower signal-to-noise ratio (SNR), varying the camera exposure time between \SIrange{300}{0.03}{\milli\second}, and with an approximate illumination overlap of \SI{67}{\percent}. 
By acquiring the calibration and evaluation data within a single scan trajectory, we ensure that the scanning position correction remains valid for the evaluation reconstructions.
In practice, calibrating the illumination field can be effectively achieved using any object that ensures high-SNR diffraction data and reliable reconstruction quality.
The binary photomask sample, the reconstructed illumination field, and the scanning pattern are shown in Supplement~1. 

\subsection{Reconstruction procedure}
In this work, we employ an ADP framework integrated with a pre-trained deep generative model, as illustrated in Fig.~\ref{setup}B.
A comprehensive description of the physics-informed ADP framework can be found in \cite{Seifert2021-fr}.
We use TensorFlow~\cite{abadi2016tensorflow} to model the optical system, leveraging its differentiable programming capabilities to seamlessly incorporate the deep generative model into our forward model. 
Specifically, the decoder of a pre-trained autoencoder serves as this deep generative model.
This enables us to represent the object as a compact latent vector rather than a conventional pixel-based image.

For each scanning position, the object is illuminated by a coherent light field. The exit field in the object plane is computed using the projection approximation~\cite{paganin2006coherent} and propagated to the detection plane via a band-limited angular spectrum method~\cite{Matsushima2009-yq}.
This yields a set of predicted diffraction patterns \(I_k\) for any given object patch and illumination field.
The optimization goal is to minimize a loss function \( L(\bm{\theta}) \) designed to maximize the likelihood that the parameter set \( \bm{\theta} \) accurately represents the observed diffraction patterns \( X_k \). The loss function is defined as follows:~\cite{seifert2023maximumlikelihood}
\begin{equation} \label{mixed_loss}
    L(\bm{\theta})=\sum_{k=1}^N\left( \ln[I_k(\bm{\theta}) + \sigma_k^2] + 
    \frac{[X_k- I_k(\bm{\theta})]^2}{I_k(\bm{\theta}) + \sigma_k^2} \right),
\end{equation}
where \( N \) is the total number of camera pixels, and \( \sigma_k^2 \) is the camera sensor readout noise determined from 300 dark measurements.
In the high-SNR calibration reconstruction, the parameter $\bm{\theta}$ encompasses the illumination field, object, object-camera distance, and the scanning positions. 
Conversely, when we assess our method's noise-robustness using latent vector reconstruction, we rely on the calibrated data for all parameters except the object. In this scenario, $\bm{\theta}$ consists solely of the latent vector representing the object.
While the optical setup and pre-calibrated illumination field are sampled at a $1024 \times 1024$ resolution, the decoder maps only to a $32 \times 32$ image output.
Hence, we employ the Mitchell-Netravali cubic filter to resize the decoder accordingly. 
This filter is chosen empirically for its high-quality output with smooth gradients and minimal aliasing artifacts~\cite{mitchell1988reconstruction}. 

All reconstructions are executed on a commercial Nvidia RTX A6000 GPU using the Adam optimizer~\cite{Kingma2014-fa} with a randomized order of diffraction patterns. The process typically completes within 100 epochs, taking approximately 20 minutes for our datasets. 
The learning rate \( \alpha \) serves as a hyperparameter that controls the step size for the gradient descent within the loss landscape. 
We find that learning rates in the range of \( \alpha = 0.1 \) to \( 1.0 \) with an exponentially decaying schedule \( \alpha_n = \alpha \times 0.97^n \) for the \( n \)-th epoch yield optimal convergence. 
While regularization terms are included for conventional reconstructions as described in \cite{seifert2023maximumlikelihood}, reconstructing the latent vectors does not necessitate any additional regularization for optimal convergence.

\subsection{Deep Generative Model}
\label{dgn}

In computational imaging, incorporating machine learning models, particularly deep generative models, offers a compelling route for enhancing image reconstruction capabilities. 
Deep generative models, such as autoencoders, are neural networks trained to learn a compressed, yet informative, representation of unlabeled data. 
This learned representation, often referred to as the latent space, captures the essential features of the data while discarding noise and redundancies. 
This section delves into the architecture, training, and characterization of the autoencoder model used in this study, elucidating how it integrates with the ptychographic reconstruction framework and with a routine to image objects of a class that is known \textit{a priori}.

\begin{figure}[!b]
\centering
\includegraphics[width=0.75\textwidth]{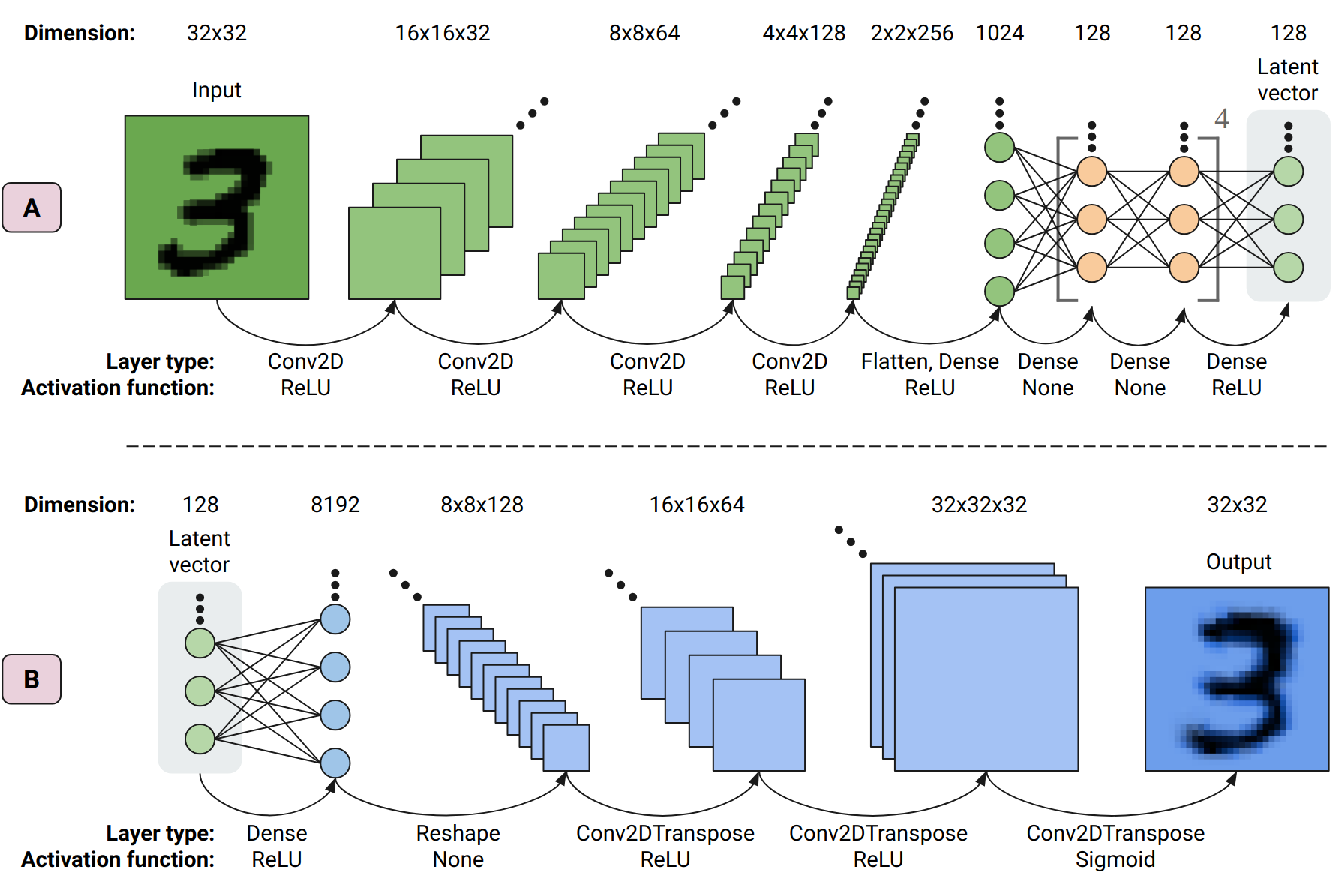}
\caption{Illustration of the autoencoder architecture, detailing output dimensions, activation functions, and layer types. (A) Schematic of the encoder model, designed to map input data into a lower-dimensional latent space. Convolutional layers employ $4 \times 4$ kernels and a stride of 2, halving the dimensions at each Conv2D layer, and are followed by rectified linear unit (ReLU) activation functions. Linear layers upstream to the latent vector facilitate rank-minimization. (B) Schematic of the decoder model, tasked with reconstructing the original input from the latent representation. Post-training on MNIST, the decoder serves as a deep generative model. The final sigmoid activation function constrains the output to the range [0, 1], making it well-suited for amplitude transmission functions in ptychography.}
\label{AE_architecture}
\end{figure}

\begin{figure}[htb]
\centering
\includegraphics[width=0.7\textwidth]{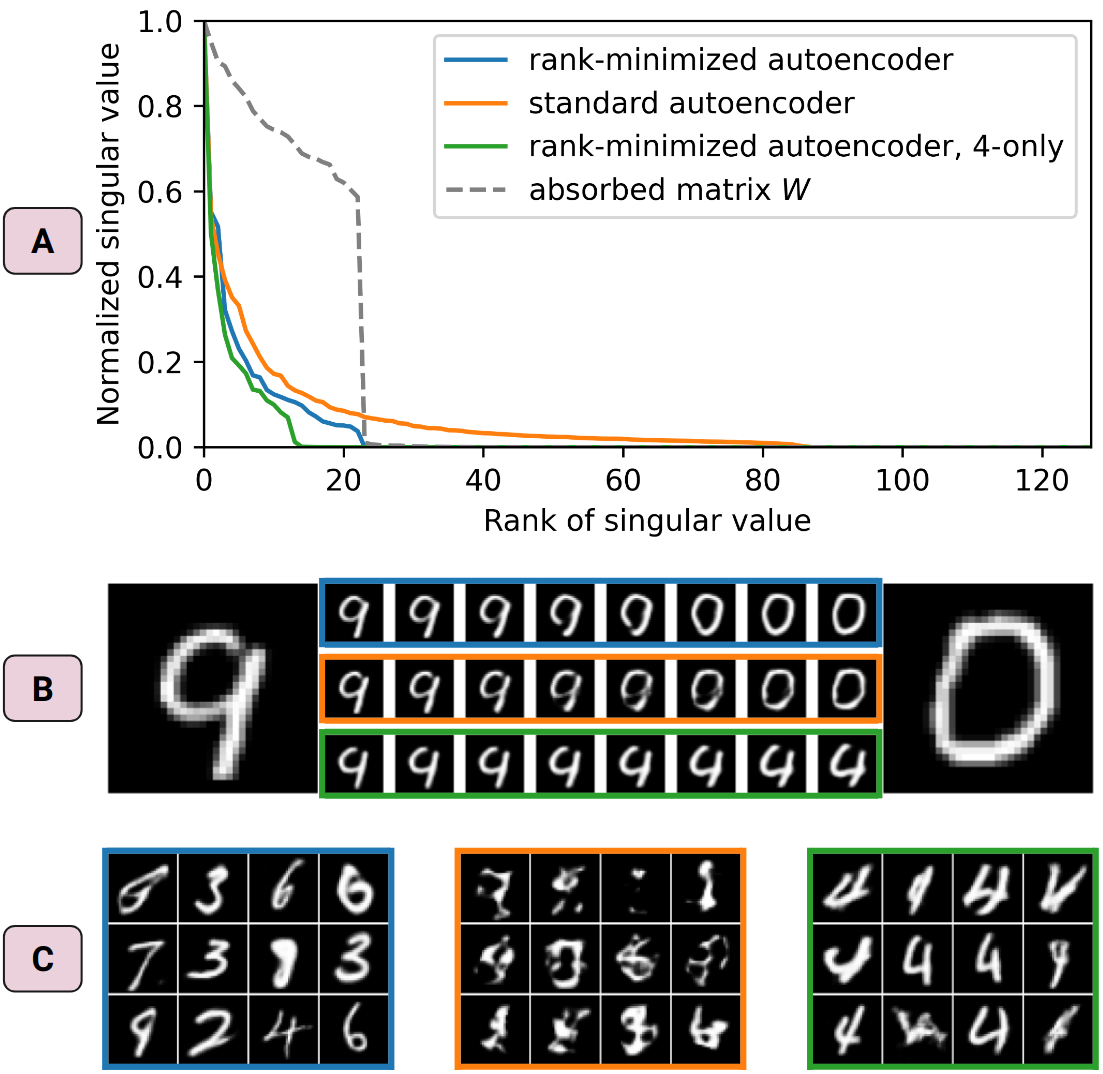}
\caption{(A) Singular values of the covariance matrix for latent vectors obtained from MNIST validation examples, revealing the effective rank of the feature representation. Different autoencoder designs and training sets are indicated by color (blue/green: implicit rank-minimized autoencoder; orange: standard autoencoder; blue/orange: full MNIST dataset; green: MNIST dataset filtered to include only '4's). (B) Linear interpolation in the latent space between two objects, illustrating the well-structured feature representation. (C) Images generated from multivariate Gaussian noise input to the decoder, highlighting the decoder's capability to produce meaningful handwritten digits. The color coding for panels (B) and (C) follows the legend provided in panel (A).}
\label{AE_characterization}
\end{figure}

The autoencoder architecture, adapted and implemented in TensorFlow from \cite{jing2020implicit}, is depicted in Fig.~\ref{AE_architecture}.
An autoencoder network typically aims to map input data into a lower-dimensional latent space and reconstruct it back to the original form~\cite{Goodfellow2016-oh}.
Mathematically, an encoder function $f$ maps an input $x$ to a latent vector $\mathbf{h} = f(x)$. A decoder function $g$ then maps $\mathbf{h}$ back to the reconstructed input $\hat{x} = g(\mathbf{h})$. 
The autoencoder is trained using the Adam optimizer and MNIST, a dataset containing 60,000 training and 10,000 validation images of handwritten digits, to minimize the binary cross-entropy loss function $L(x, \hat{x} = g(f(x)))$.
The training is a one-time process and requires only 50 epochs which take about 20 minutes on an Nvidia RTX A6000 GPU.
With respect to the image size of our object, the design of our under-complete autoencoder requires a latent space of significantly smaller dimensionality. 
Otherwise, the autoencoder would trivially learn the identity function, failing to capture the most salient features of the training data.
This raises the question of selecting the optimal latent space dimension.
Rather than relying on a heuristic trial-and-error approach, we employ the Implicit Rank-Minimizing Autoencoder (IRMAE) model~\cite{jing2020implicit}.
The IRMAE includes eight additional linear layers $W_1, W_2, ..., W_8$ at the end of the encoder network, which are randomly initialized.
Deep linear networks have been shown to induce implicit regularization, leading to low-rank solutions~\cite{NEURIPS2019_c0c783b5}.
Hence, we can choose the latent dimension of $\mathbf{h}$ to be reasonably large (128 in our case) and let the training process automatically find the lowest rank.

In Fig.~\ref{AE_characterization}, we evaluate different autoencoder architectures across various scenarios.
Fig.~\ref{AE_characterization}A illustrates the singular value decomposition on the covariance matrix of MNIST validation examples to assess the effective rank needed for feature representation.
Our findings indicate that the rank-minimized autoencoder has an effective rank of 22.
This is corroborated by the same rank of the matrix absorbed into the encoder, calculated as $W = \prod_{i=1}^{8} W_i$.
In contrast, we find that the singular values for a standard autoencoder, where $W$ is an identity matrix, can only be neglected above a rank of 86.
Hence, the rank-minimized autoencoder spans a more compact latent space.
Additionally, we examine the impact of simplifying the training set by using only the 982 handwritten '4's from the MNIST dataset, which results in a lower rank of 13.
This adjustment in training samples can be viewed as imposing a stronger prior belief about the object in the context of latent vector reconstruction in computational imaging. 

In Fig.~\ref{AE_characterization}B, we further explore the properties of the latent space. 
Specifically, we illustrate latent vector arithmetic through linear interpolation between two latent vectors representing a '9' and a '0'. 
Using the rank-minimized autoencoder, we observe that the latent space interpolation captures smooth and meaningful transitions between different images, indicating a well-structured feature representation.
However, this is less evident in the standard autoencoder, which results in a more ambiguous interpolation between images.
In the scenario where the model is trained only on a subset of the MNIST dataset, images that are not part of the training set are not as accurately represented. 
This is by design, as the deep generative model can be intentionally optimized to reconstruct objects within its trained class, in this case, the digit '4'. 
This targeted approach offers an advantage for applications where more concrete prior knowledge about the object class is available.

Lastly, we sample images generated from multivariate Gaussian noise input to the decoder in Fig.~\ref{AE_characterization}C. 
Remarkably, images generated using rank-minimized decoders predominantly resemble handwritten digits rather than random patterns that are observed with the standard decoder.
This is particularly advantageous for our ptychography application, as it implies that the reconstruction process is intrinsically guided towards generating images that align with prior knowledge -- here, the class of handwritten digits. 
We seek out this property to enhance the robustness of our method, especially when dealing with ill-posed data, by reducing the likelihood of spurious reconstructions.
An extended evaluation of the autoencoder's performances across various inputs is available in Supplement~1.

\section{Results}
\subsection{Experimental Amplitude Transmission Reconstructions}

\begin{figure}[htb]
\centering
\includegraphics[width=0.74\textwidth]{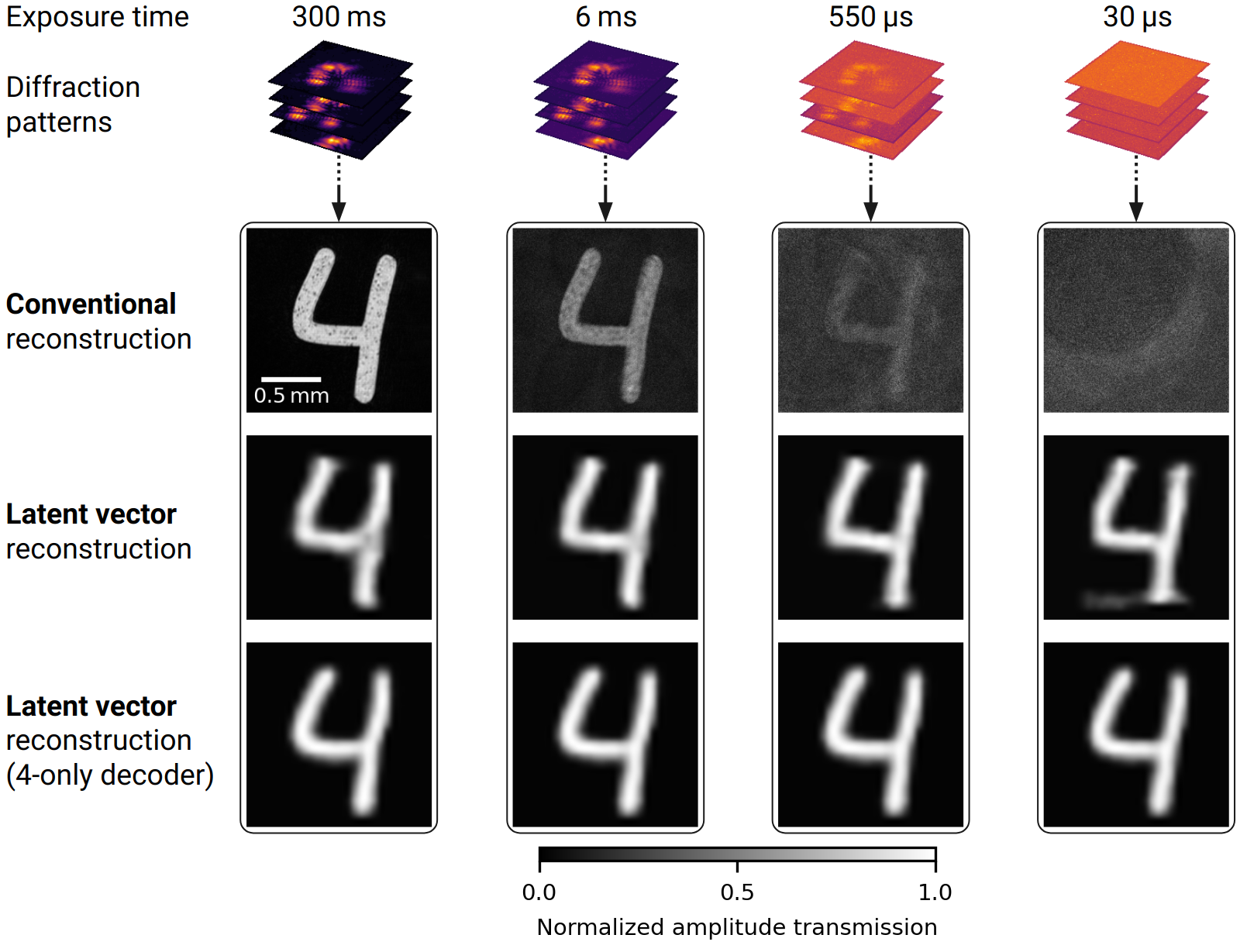}
\caption{Comparison of ptychographic amplitude image reconstruction results under varying signal-to-noise ratios (SNR). The top row displays stacks of diffraction patterns used for reconstruction, with exposure times decreasing from left to right, leading to a corresponding decrease in SNR. The second row presents results from conventional reconstruction. Subsequent rows feature latent vector reconstructions using a pre-trained deep generative model, first trained on the full MNIST dataset and secondly on a filtered MNIST dataset containing only images resembling the digit '4'.}
\label{reconstruction_results}
\end{figure}

We present the main experimental result of this paper in Fig.~\ref{reconstruction_results}. 
We illuminate an amplitude-only sample shaped like the digit '4' and adjust the camera's exposure time over four orders of magnitude. 
As a result, we acquire sets of diffraction patterns ranging from high SNR to extremely low SNR.
These datasets are then used for ptychographic reconstructions.
In conventional reconstruction, using a $468 \times 468$ pixel basis, the object's amplitude transmission function is pristine at the highest SNR but deteriorates to noise at a \SI{30}{\micro\second} exposure. 
In contrast, switching to latent space reconstruction with a pre-trained deep generative model significantly improves low-SNR performance. The reduced rank of the latent space, previously shown to be 22, cuts the number of free parameters by approximately $10,000\times$ compared to the conventional reconstruction method. This allows for successful object determination with remarkably fewer photons. Moreover, the model trained specifically on the digit '4' shows more stable convergence and slightly better reconstructions overall.

In the high-SNR scenario, the conventional reconstruction outperforms our latent vector approach in terms of image sharpness. 
When a sufficient number of detected photons is available, reducing the number of parameters that represent the object offers no advantage.
Indeed, this becomes a drawback when the lower-resolution output of our deep generative model is resized to match the higher resolution used in our ADP framework, resulting in limited image sharpness.
This can be interpreted as a trade-off for the immense reduction of free parameters.

During the optimization process, we observe that the latent vector reconstruction based on training with the full MNIST dataset can occasionally get stuck in local minima, even when using diffraction data with high SNR.
Due to the randomization of the order of the diffraction patterns in the stochastic gradient descent, the first update step of the latent vector can walk in a different direction within the loss function optimization landscape.
The latent space is non-injective, meaning that multiple different latent vectors can map onto the same output. 
Therefore, the optimization procedure is sensitive to the initial state and first gradient step in particular.
In practice, we find that this sensitivity can be mitigated by initializing the latent vector as the vector average $\bar{\mathbf{h}}$ obtained using the pre-trained encoder function $f(x)$ and all training examples $x_{\text{train}}$ expressed as
\begin{equation}
    \bar{\mathbf{h}} = \frac{1}{N_x} \sum_{i=1}^{N_x} f(x_{\text{train},i}),
\end{equation}
where $N_x$ is the total number of training samples, and $f(x_{\text{train},i})$ is the latent vector corresponding to the $i^{th}$ training sample.
This initialization approach is practical as the training samples are readily available from the pre-training phase.
Moreover, for the case of the deep generative model trained on the filtered MNIST dataset only including '4's, this vector average initialization is not required. 
In this case, the convergence is consistently stable and fast, and the latent vector can be initialized as random or uniform.

\subsection{Numerical Reconstructions and Quantitative Comparisons}

To quantitatively assess the noise robustness of our method and its ability to generalize to another object, we simulate ptychographic reconstructions using a known amplitude-only hand-drawn '3' as ground truth. We generate diffraction patterns using the probe and scanning pattern shown in Supplement~1, with an object-camera distance of \SI{8}{cm}, while all other simulation parameters are consistent with the experimental setup. We vary the total photon count in the illumination field from \(10\) to \(10^6\)\,photons, assuming a uniform camera readout noise of $\sigma_k = \SI{0.3}{photons}, \forall k$.

\begin{figure}[b]
\centering
\includegraphics[width=0.9\textwidth]{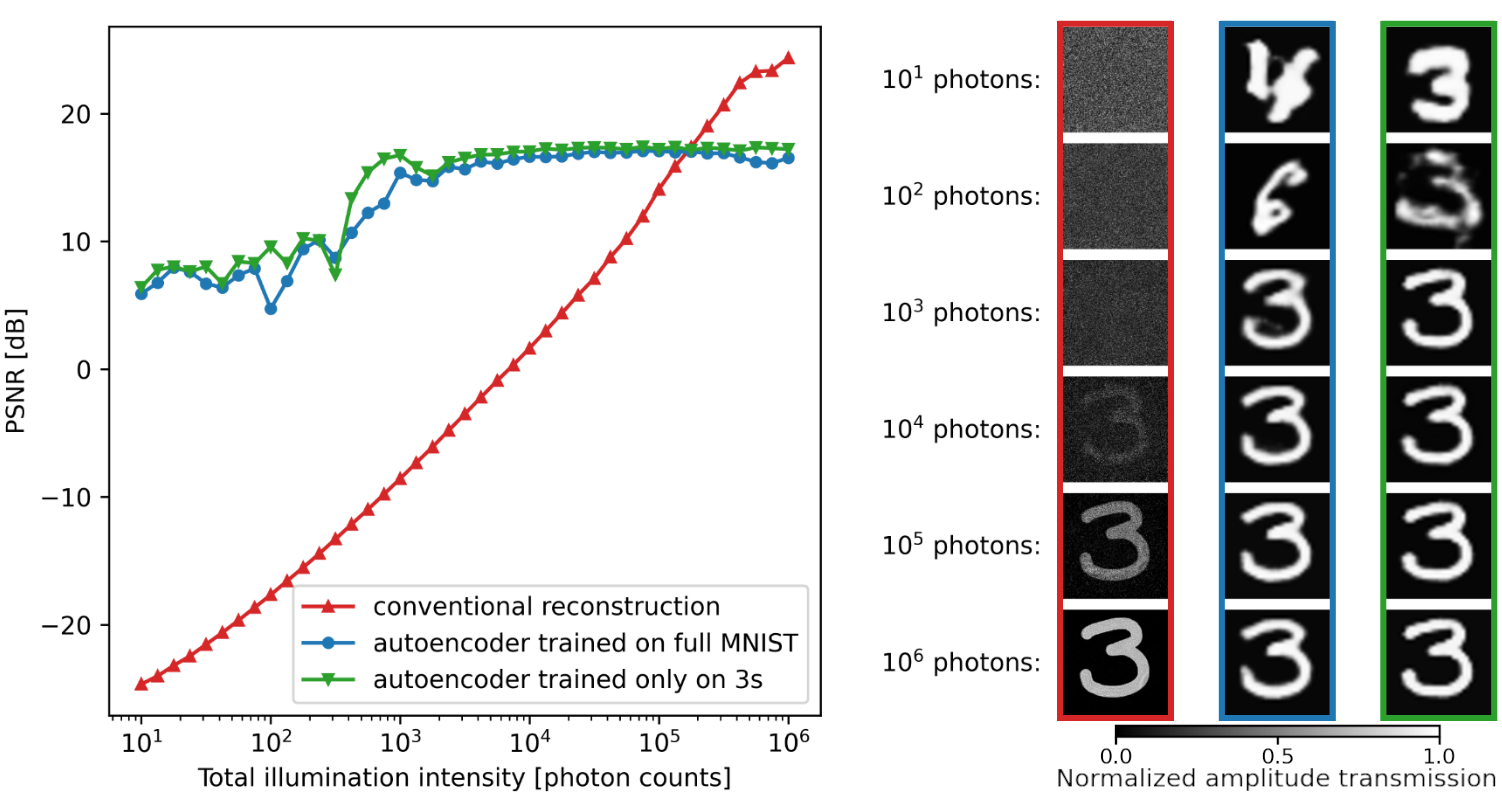}
\caption{Comparison of reconstruction quality for conventional and latent vector ptychographic methods across varying signal-to-noise ratios (SNRs), obtained from numerical simulations. The Peak Signal-to-Noise Ratio (PSNR) serves as the quality metric and is plotted against the total photon count in the illumination field. Selected object amplitude reconstructions are displayed for various photon counts to highlight the trade-offs between the methods. A latent vector reconstruction using a deep generative model trained on filtered data is also included for comparison.}
\label{simulation_results}
\end{figure}

To evaluate reconstruction quality in Fig.~\ref{simulation_results}, we use the Peak Signal-to-Noise Ratio (PSNR) between the reconstructed image $\hat{x}$ and the ground-truth $\hat{x}_{\text{gt}}$. As the maximum pixel value for our decoder output is equal to one, we can write 
\begin{equation}
    \text{PSNR} = -10 \log_{10}\left(\frac{1}{N} \sum_{i=1}^{N} (\hat{x}_i - \hat{x}_{\text{gt}, i})^2\right).
\end{equation}
In the double-logarithmic plot, a linear relationship between illumination intensity and conventional reconstruction quality suggests a power-law behavior. The curve plateaus at \(10^6\)\,photons, indicating the inverse problem becomes well-posed. This is further supported by the amplitude reconstruction visually nearing its optimum in sharpness and contrast at the highest simulated SNR.

For the latent vector reconstruction, successful convergence occurs at around \(10^3\)\,photons for both the full and filtered MNIST-trained models. This corresponds to an average photon count of 0.001 per camera pixel. Below this threshold, the PSNR is inflated due to the deep generative model's inability to generate noisy outputs matching the readout noise, leading to spurious correlations with the ground truth. The model trained exclusively on '3's demonstrates stable convergence at marginally lower SNRs, specifically around a few hundred photons in the illumination field.

In summary, our method excels at low SNRs where the conventional method struggles to produce meaningful reconstructions, although it is inherently limited at high SNRs due to the reduced degrees of freedom in the deep generative model. Hence, when sufficient information is available in the diffraction data, the conventional reconstruction method should be preferred.

\subsection{Loss Landscapes}

\begin{figure}[b]
\centering
\includegraphics[width=0.8\textwidth]{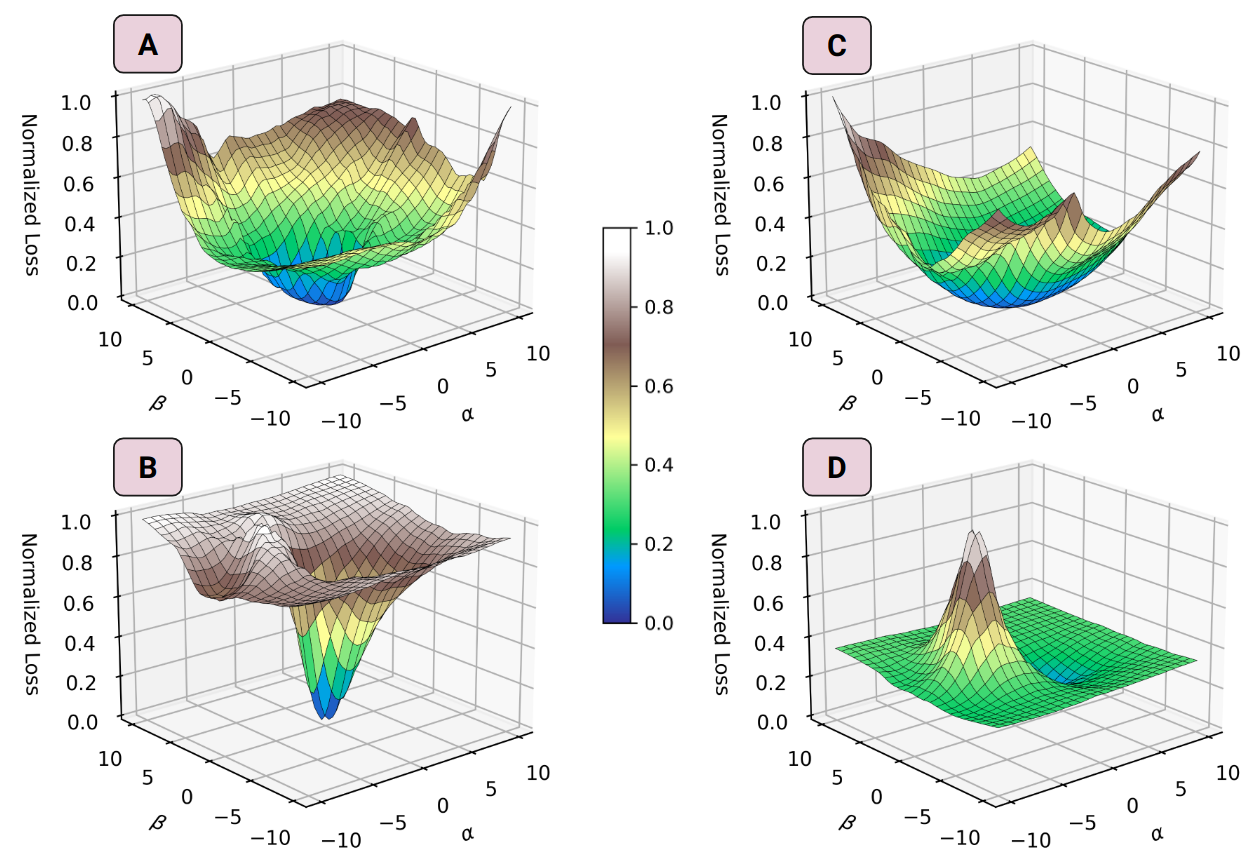}
\caption{Visualization of the optimization loss landscapes for different scenarios. $\alpha$ and $\beta$ are coefficients for the two leading principal components of the latent space used for ptychographic reconstruction. (A) The landscape for high signal-to-noise ratio (SNR) and training on the full MNIST dataset. (B) The landscape when reconstructing from low-SNR (high-noise) diffraction data. (C) The landscape after training the deep generative model on a filtered MNIST dataset containing only the digit '4'. (D) The landscape when optimizing using a Poisson-only loss function at high SNR, for comparison with the mixed Poisson-Gaussian loss from all other panels.}
\label{landscapes}
\end{figure}

Given the compact nature of the latent space, we have the unique opportunity to approximate and visualize the loss landscape that is traversed during optimization (see Fig.~\ref{landscapes}). Utilizing the two leading orthogonal principal components of the latent space, we construct a three-dimensional representation of the loss landscape.
We perform a principal component analysis on the covariance matrix of all latent vectors from the validation MNIST dataset to identify the two most informative directions associated with the two leading singular values, denoted as $\mathbf{v}_1$ and $\mathbf{v}_2$. 
Given an optimal latent vector $\mathbf{h}_{\text{opt}}$ obtained from experimental diffraction data, we explore the loss landscape by varying this optimal point along the directions of $\mathbf{v}_1$ and $\mathbf{v}_2$:
\begin{equation}
\label{pca_eq}
    \mathbf{h}(\alpha, \beta) = \mathbf{h}_{\text{opt}} + \alpha \mathbf{v}_1 + \beta \mathbf{v}_2.
\end{equation}
Here, $\alpha$ and $\beta$ range from -10 to 10 based on the extent of these variations in image space as illustrated in Supplement~1. 
We then compute the loss value for each $\mathbf{h}(\alpha, \beta)$ to visualize the landscape.

In the high-SNR case with training on the full MNIST dataset (Fig.~\ref{landscapes}A), the loss landscape exhibits a distinct but non-convex and asymmetric minimum. This topography accounts for the sensitivity to the initial latent vector state. The non-convexity is exacerbated in low-SNR scenarios (Fig.~\ref{landscapes}B), confirming that convergence is more challenging when the signal is weak. 
Interestingly, the loss landscape exhibits a smoother and more convex topography when the model is trained exclusively on images resembling the digit '4' (see Fig.~\ref{landscapes}C). This finding aligns with our previous observation that such specialized training renders the optimization process less prone to getting trapped in local minima and more forgiving of suboptimal latent vector initialization.
Finally, we explore the impact of using an alternative loss function based solely on Poisson statistics in Fig.~\ref{landscapes}D. This loss function, defined as $L_\mathrm{P}(\bm{\theta})=\sum_{k=1}^N\left( \sqrt{X_k}- \sqrt{I_k(\bm{\theta})}\right)^2$, reveals a loss landscape with large plateaus and a higher degree of non-convexity for the high-SNR, full MNIST-trained scenario.
These characteristics align with our observation that the mixed Poisson-Gaussian loss function generally offers better convergence behavior in comparison.

\section{Discussion and Conclusion}

We present a novel approach to ptychographic image reconstruction by integrating a deep generative model into a physics-informed Automatic Differentiation Ptychography (ADP) framework.
By incorporating prior knowledge about the object class of a specimen, our method significantly reduces the number of free parameters in the optimization problem, enabling robust reconstructions in low signal-to-noise ratio (SNR) scenarios. 
Since the pre-training of the generative model and the ADP reconstruction are separated, this approach is modular and portable to related AD-based imaging methods. 
For instance, the deep generative model can seamlessly be exchanged or reused in an adapted physics-informed imaging modality without retraining. 
As the field of generative artificial intelligence is currently undergoing an immense research interest, our work presents a straightforward way to incorporate future latent generative models into computational imaging.

One key observation of this paper is the inherent trade-off between noise robustness and the maximum achievable fidelity of the reconstructed image. This limitation is primarily due to the output resolution constraints of the pre-trained decoder. 
This opens up intriguing avenues for future research, including the exploration of alternative deep generative models for ptychographic reconstruction. While Generative Adversarial Networks (GANs) and latent diffusion models have shown promise in related imaging contexts such as compressed superresolution imaging through multimode fibers~\cite{Li2022-on} or high-resolution image reconstruction from human brain activity~\cite{Takagi2022.11.18.517004}, they introduce their own set of challenges. 
These include increased computational complexity, less interpretable latent spaces~\cite{asperti2023comparing}, and the added model complexity required for capturing high-resolution or complex greyscale features, which could compromise our method's ability to reconstruct from severely ill-posed data.
Indeed, architectures like the recently shown GigaGAN~\cite{kang2023gigagan} show excellent generative abilities and a controllable latent space, but require up to 4700 days of training on a high-end A100 GPU, while the autoencoder design in this work trains within minutes on any commercial laptop.

Another promising direction for future research in latent vector reconstruction is the extension of the model to output complex-valued images. This would lift the current limitation of representing only amplitude objects, thereby broadening the applicability of our approach. Cherukara~\textit{et al.} have already made strides in this direction, demonstrating ptychographic imaging with a Y-shaped latent model that performs an end-to-end mapping from diffraction patterns to both phase and amplitude images~\cite{cherukara2020real}. This is particularly interesting for noise-robust latent vector reconstruction, as the shared latent representation could be leveraged to account for the high correlation typically observed between phase and amplitude in complex objects, akin to the implementation recently shown in~\cite{Chang2023-yk}. Consequently, even though the output dimensionality would effectively double to accommodate both phase and amplitude, the rank of the latent space may not necessarily need to double, thanks to this inherent correlation.

Our utilization of a compact latent space for object representation provides a unique opportunity to approximate and visualize the optimization loss landscape during ptychographic object retrieval. 
This offers valuable insights into the convergence behavior and the sensitivity to the initial state of the reconstruction process. 
However, the mapping from the latent space to the object space is non-injective, and our landscape visualization, therefore, is localized and truncated due to the omission of principal components associated with smaller but still relevant eigenvalues.

Our method's robustness in low-SNR conditions offers valuable applications in both biological and industrial settings where prior knowledge for pre-training is often available. 
It is particularly well-suited for medical imaging of delicate specimens, where minimizing radiation dose is a priority. 
The reduced computational complexity, thanks to fewer free parameters and the calibrated illumination field, also makes it ideal for real-time and specialized imaging scenarios like industrial quality control, where quick yet quality-assured reconstructions are desired.
Indeed, we occasionally observe high-SNR latent vector reconstructions to converge within a single epoch, as compared to the multiple epochs typically required for conventional reconstructions. 
Furthermore, our method's noise resilience makes it a promising tool for imaging in photon-starved regimes, such as extreme ultraviolet (EUV) or X-ray applications, where imaging with minimal photon counts is often required.

In conclusion, our work represents a step forward in the field of computational imaging by demonstrating the power of integrating machine learning techniques with physics-based models for robust image reconstruction.
As computational imaging continues to evolve, the integration of deep learning models with traditional imaging techniques promises to unlock new capabilities and applications across a wide range of scientific and industrial domains. \\

\noindent \textbf{Funding:} Netherlands Organization for Scientific Research NWO (Perspective P16-08). \\

\noindent \textbf{Disclosures:} The authors declare no conflicts of interest. \\

\noindent \textbf{Acknowledgments:} We thank Dorian Bouchet for helpful discussions and Cees de Kok, Dante Killian, Jan Bonne Aans, Aron Opheij, Paul Jurrius and Arjan Driessen for technical support.  \\

\noindent \textbf{Data availability:} Experimental raw data, synthetic data, and code underlying these results are available in~\cite{Seifert2023-bk} under open licenses. This includes the camera readout noise measurements as well as the ptychographic calibration and validation datasets.
Using the provided Python and TensorFlow code, a deep generative model can be trained and used to perform latent vector reconstruction within a ptychography reconstruction framework based on automatic differentiation.\\

\noindent \textbf{Supplemental Document:}
See Supplement 1 below for supporting content.


\section*{REFERENCES}

\bibliography{bibliography}

\bigskip



\onecolumngrid
\clearpage
\beginsupplement
\begin{center}
\textbf{\LARGE Noise-robust latent vector reconstruction in ptychography using deep generative models: Supplement 1}

\bigskip
Jacob Seifert,$^1$, Yifeng Shao,$^{2}$ and Allard P. Mosk$^1$\\ \vspace{0.15cm}
\textit{\small $^\mathit{1}$Nanophotonics, Debye Institute for Nanomaterials Science and Centre for Extreme Matter and Emergent Phenomena, Utrecht University, P.O. Box 80000, 3508 TA Utrecht, The Netherlands}\\
\textit{\small $^\mathit{2}$Imaging Physics Department, Applied Science Faculty, Delft University of Technology, The Netherlands}\\
\end{center}
\vspace{1cm}

This document provides supplementary material to \textit{Noise-robust latent vector reconstruction in ptychography using deep generative models}.

\section{Autoencoder performance across varied inputs}

\begin{figure}[tbh]
\centering
\includegraphics[width=\textwidth]{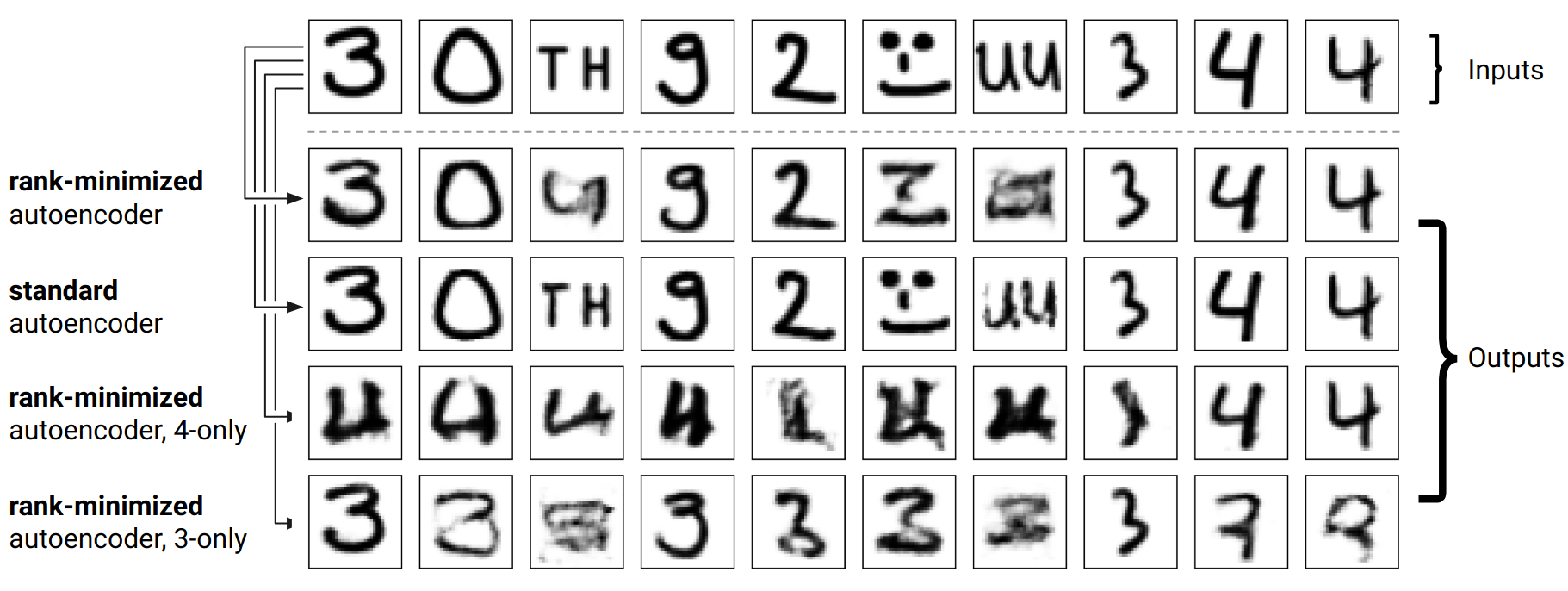}
\caption{Performance comparison of various autoencoder architectures using ten hand-drawn images as inputs. The figure illustrates the capability of each model to faithfully reconstruct or generalize to these images, highlighting the trade-offs between standard and rank-minimized autoencoders, and the utilization of filtered training data.}
\label{AE_overview}
\end{figure}

As elaborated in Section~2C of the main document, we explore various autoencoder architectures. In this section, we extend that discussion by examining how these architectures map different types of inputs to outputs. Fig.~\ref{AE_overview} provides a comprehensive view of the autoencoder's performance on a range of inputs, including both hand-drawn digits and objects not present in the training set. 

The standard autoencoder, with its larger rank of 86, offers the most faithful input-to-output mapping. Notably, it can generalize well to unseen objects like alphabets and even a hand-drawn smiley. However, in the context of this work, we are primarily interested in deep generative models that can efficiently map to hand-drawn digits, as we have prior knowledge that the objects of interest belong to this specific class. In this regard, the implicit rank-minimized autoencoder proves to be more suitable as it spans a more compact latent space, which is beneficial for the purpose of reconstructing images from ill-posed data.

Additionally, we investigate the impact of training the autoencoder on a filtered dataset, focusing on specific digits like '3' and '4'. We find that such specialized training only maintains the mapping for those specific digits and also tends to "morph" other inputs into resembling these digits. This demonstrates the flexibility in incorporating varying degrees of prior knowledge about the object under study. In our application, this could range from knowing that the object is \textit{any} hand-drawn digit to knowing that it is a \textit{specific} digit.

\section{Principle component analysis of the latent space}

To explore the latent space of the trained autoencoder, we aim to identify the two leading principal components. These components will serve as directions along which the latent vectors can be varied to generate visualizations of the loss landscape.
First, we encode the set of 10,000 validation MNIST images $x_\text{test}$ using the trained encoder function, obtaining their latent representations $\mathbf{h_\text{test}} = f(x_{\text{test}})$. Next, we calculate the covariance matrix $\mathbf{C}$ of the latent vectors as
\begin{equation}
\mathbf{C} = \text{cov}(\mathbf{h_\text{test}}).
\end{equation}
We then perform singular value decomposition (SVD) on the covariance matrix $\mathbf{C}$:
\begin{equation}
\mathbf{U}, \bm{\Sigma}, \mathbf{V} = \text{SVD}(\mathbf{C}).
\end{equation}
Here, $\mathbf{V}$ contains the eigenvectors of $\mathbf{C}$, sorted by their corresponding eigenvalues in descending order.
The principal components corresponding to the two largest eigenvalues are then the first and second rows of $\mathbf{V}$, denoted as $\mathbf{v}_1$ and $\mathbf{v}_2$, respectively.

\begin{figure}[tbh]
\centering
\includegraphics[width=\textwidth]{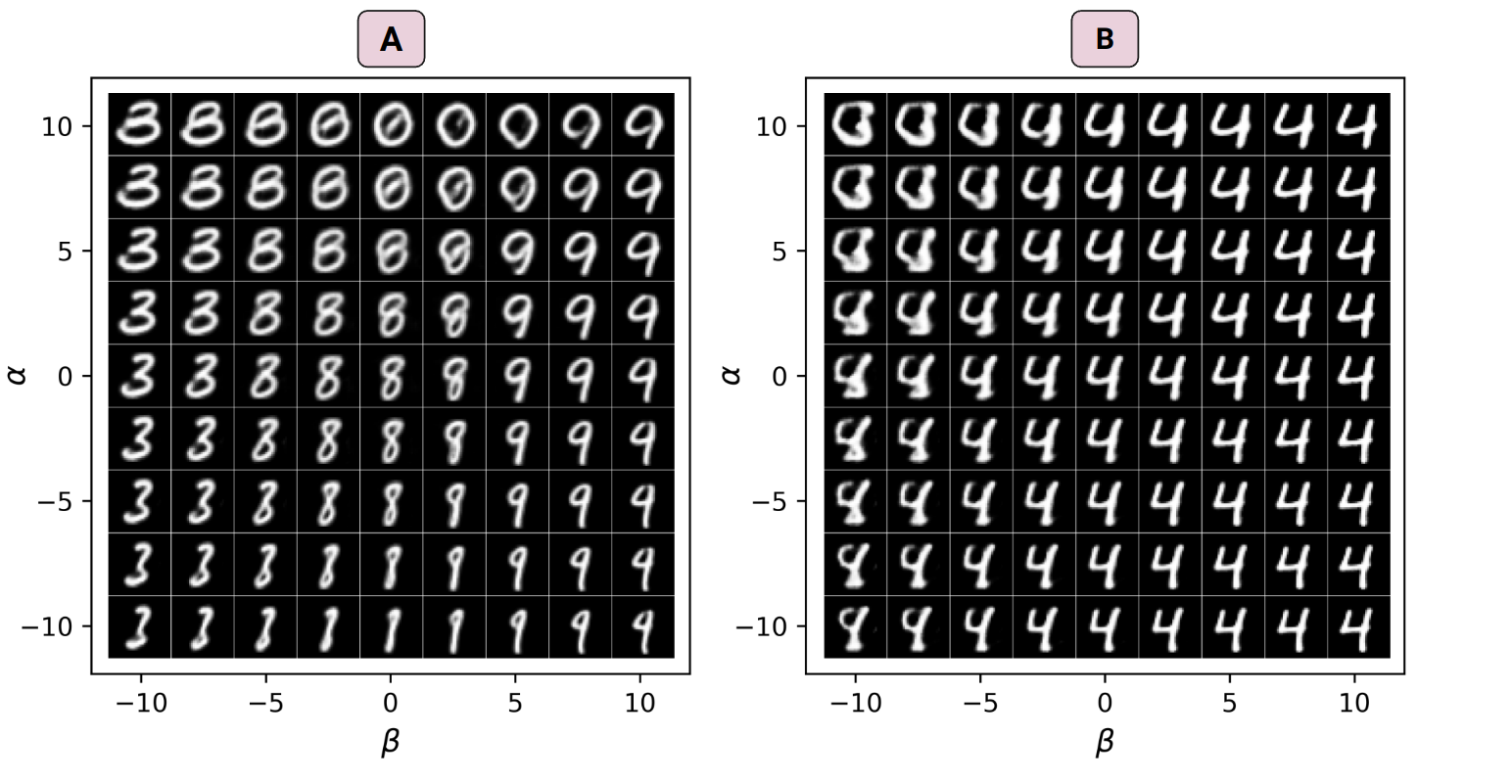}
\caption{Exploration of the latent space along the most informative principal components. (A) Generated images obtained by varying the average latent vector along the two leading principal components. (B) Generated images obtained by varying a latent vector representing the digit '4' along the same principal components.}
\label{eigenvectors}
\end{figure}

To generate new images along the directions of these two orthogonal leading directions, we interpolate a new latent vector $\mathbf{h}$ as described in Eq.~4 in the main document, where $\alpha$ and $\beta$ are scalar values that define the extent to which the latent vector is varied along each principal component. 
By following this procedure, we can explore the latent space along its most informative directions, thereby generating meaningful variations of the original images as illustrated for the average latent vector in Fig.~\ref{eigenvectors}A and a latent vector representing a '4' in Fig.~\ref{eigenvectors}B.

\section{Sample calibration and scanning patterns}

\begin{figure}[htb]
\centering
\includegraphics[width=\textwidth]{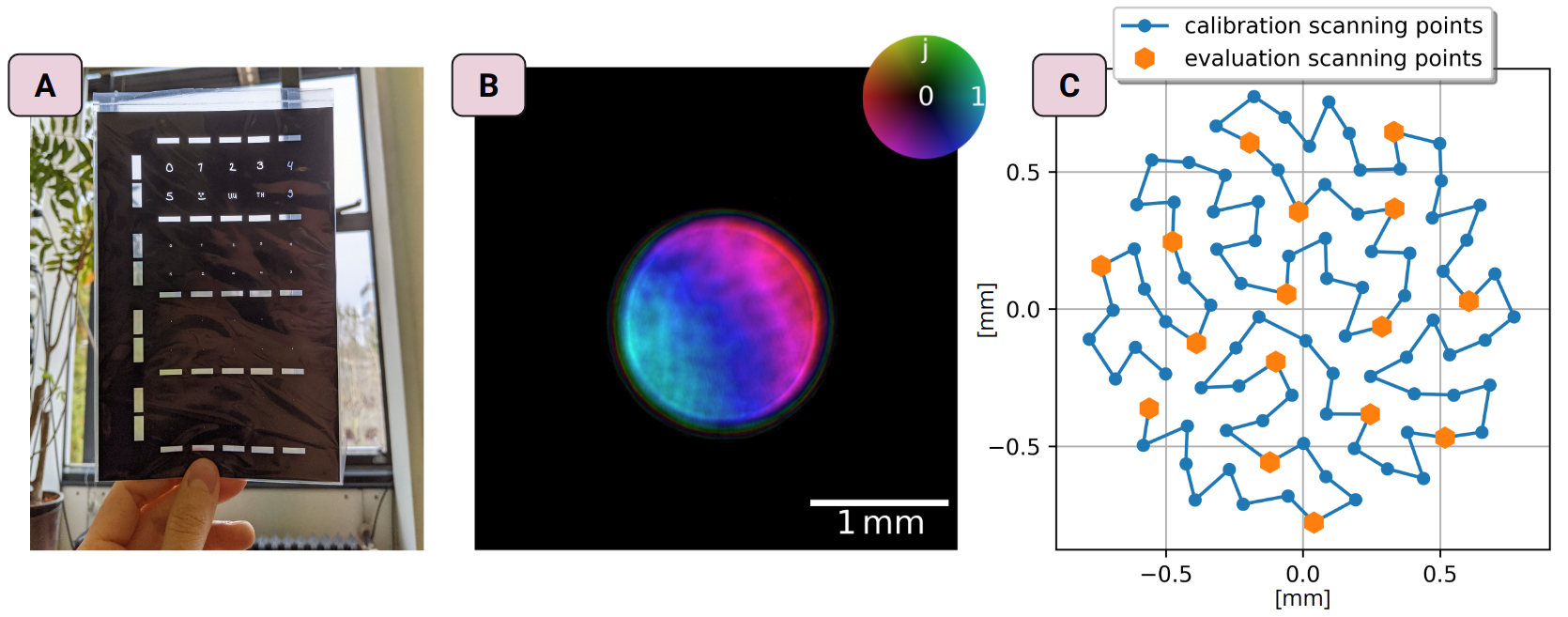}
\caption{(A) Photograph of the photomask used as the binary amplitude object. The object consists of rows of hand-drawn binary shapes, each row scaled by a factor of three. The '4' in the third row, which is used in this work, has an object size of approximately \SI{1}{mm^2}. (B) Calibrated illumination field, obtained through a high-overlap ptychographic calibration reconstruction. The total photon count is approximately \SI{950e6}{photons}. (C) Scanning pattern used for calibration and noise-robustness evaluation reconstructions.}
\label{calibration}
\end{figure}

In this section, we present key experimental parameters employed in both experimental and numerical studies. 
Fig.~\ref{calibration}A showcases a photograph of the photomask that serves as the binary amplitude object. This photomask features rows of binary hand-drawn objects akin to those previously depicted in Fig.~\ref{AE_overview}. Each row is successively scaled down by a factor of three with respect to the previous row. Specifically, we focus on the '4' in the third row, which has an object size of approximately \SI{1}{mm^2}.
Fig.~\ref{calibration}B and C display the calibrated illumination field and the scanning pattern, respectively. The scanning pattern is a high-overlap trajectory comprising 96 points arranged in a Fermat spiral shape, used solely for calibrating the illumination field, scanning points, and object-camera distance. To assess the noise-robustness of our latent vector reconstruction approach, we utilize only the 16 orange, uniformly distributed scanning points within this pattern.

\begin{figure}[bth]
\centering
\includegraphics[width=\textwidth]{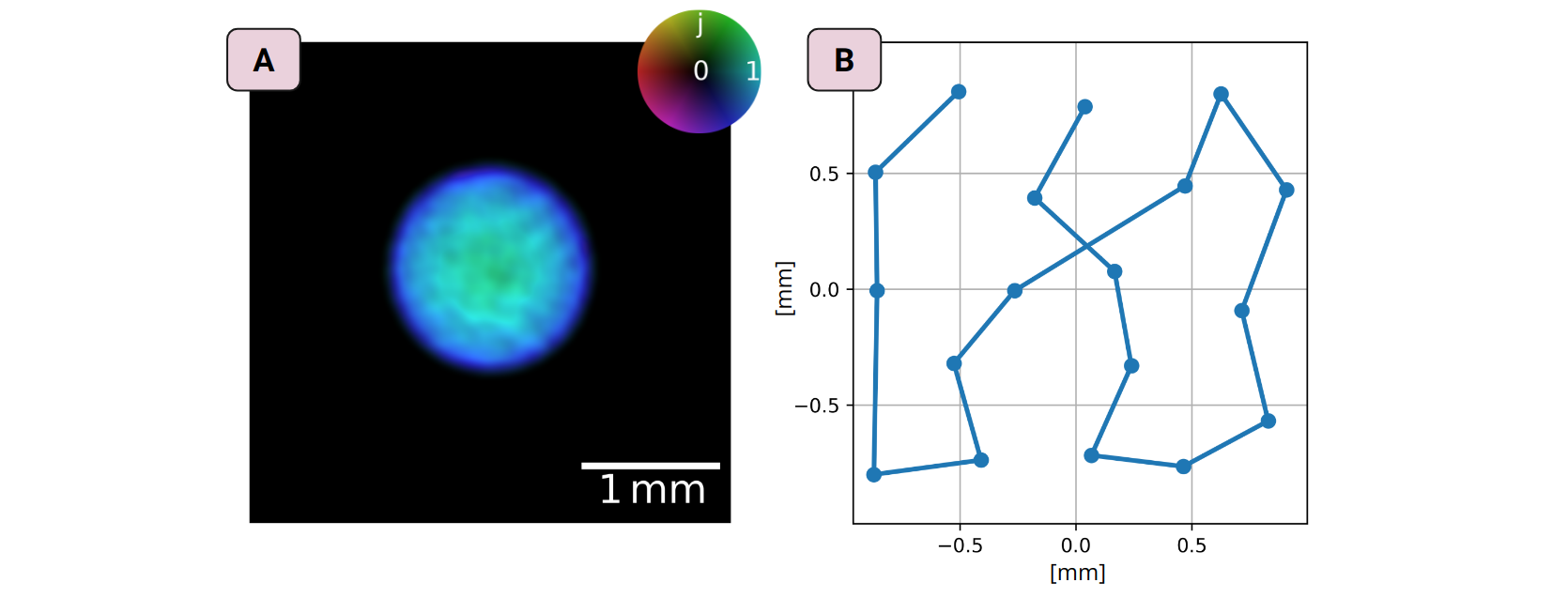}
\caption{(A) Illumination field used in the numerical simulations. (B) Scanning pattern generated through Poisson disk sampling.}
\label{simulation_probe}
\end{figure}

Finally, Fig.~\ref{simulation_probe} illustrates the normalized illumination field and scanning pattern utilized in our numerical simulations. In this case, we adopt Poisson disk sampling to generate the scanning trajectory. Much like the Fermat spiral, this choice is well-suited for ptychography as it yields an arbitrary, non-regular grid while maintaining uniform distances between neighboring points~\cite{Bridson2007-mk}.
In these simulations, the magnitudes of the illumination fields are rescaled to achieve a total photon count ranging from $10$ to $10^6$\,photons.

\end{document}